\documentstyle[prl,aps,epsf]{revtex}         
\tighten

\begin{document}
\twocolumn[\hsize\textwidth\columnwidth\hsize\csname@twocolumnfalse\endcsname

\title{Does matter wave amplification work for fermions?}
\author{Wolfgang Ketterle and Shin Inouye}
\address{Department of Physics and Research Laboratory of Electronics, \\
Massachusetts Institute of Technology, Cambridge, MA 02139}

\date{\today }
\maketitle
\begin{abstract}
We discuss the relationship between bosonic stimulation, density fluctuations, and matter wave
gratings.  It is shown that enhanced stimulated scattering, matter wave amplification and atomic
four-wave mixing are in principle possible for fermionic or non-degenerate samples if they are
prepared in a cooperative state.  In practice, there are limitations by short coherence times.
\end{abstract}
\pacs{PACS numbers: ?? 03.75.Fi,34.50.-s,67.90.+z} \vskip1pc ]

{\it Introduction}. The realization of Bose-Einstein condensation in atoms has made it possible to
study the phenomenon of bosonic stimulation for massive particles.  Superradiance of atoms
\cite{inou99super}, four wave mixing \cite{deng99} and matter wave amplification
\cite{inou99mwa,kozu99amp} were described as processes which are bosonically stimulated, i.e.,
their rates are proportional to $(N+1)$, where $N$ is the number of identical bosons in the final
state. These experimental achievements have raised the question whether these processes are
inherently connected to bosonic systems.

We have recently pointed out that atomic superradiance does not depend on Bose-Einstein statistics
and would occur for thermal atoms or even for fermions, although with a much shorter coherence
time \cite{inou99super}, and similar arguments should apply to four-wave mixing.  These
suggestions have stirred a controversy among researchers.  This note will reconcile the different
physical descriptions.  The central result is that the stimulated processes mentioned above do not
rely on quantum statistics, but rather on symmetry and coherence.

This note also addresses a widespread misconception about bosonic stimulation which regards
stimulated scattering as being solely due to quantum-statistical enhancement by the final state,
i.e. as if the particles in the final state mysteriously attract other identical particles
{\it without any other physical effect}.  We show that the presence of a macroscopically occupied
state increases the density fluctuations of the system, and bosonically enhanced scattering is
simply the diffraction of particles from these density fluctuations.

{\it Scattering theory}. It is useful to summarize basic aspects of the theory of scattering of
light or particles from an arbitrary system.  These textbook results simply follow from lowest
order perturbation theory (Fermi's Golden Rule). The double differential cross section can be
decomposed into two factors $\frac{{\rm d}^2\sigma}{{\rm d}\Omega \: {\rm d}\omega}= \left(
\frac{{\rm d}\sigma}{{\rm d}\Omega} \right)_{\rm single} S(q,\omega)$.  The first one is the
differential cross section for the scattering by a single particle (e.g. the Rayleigh cross
section for far-off resonant light scattering), the second one is the dynamic structure factor
(van Hove or scattering function) $S(q, \omega)$ which is the Fourier transform of the
density-density correlation function: $ S(q,\omega)=(1/2\pi) \int{{\rm d}t \: e^{{\rm i}\omega t}
\langle \hat\rho(q,t) \hat\rho^\dag(q,0)} \rangle $ where $\hat\rho(q)$ is the Fourier transform
of the particle density operator (see, e.g.\ \cite{grif93}).

For a non-interacting system, $S(q,\omega)$ can be expressed using the single-particle states
$|i\rangle$ with energy $E_i$ and occupation numbers $N_i$:

\begin{eqnarray}
\lefteqn{ S(q,\omega ) = S_0(q) \delta(\omega) +}  \nonumber \hspace{0.15 in} \\
&& \sum\limits_{i \neq j}  \left| {\left\langle j \right|e^{{\rm i}qr} \left| i \right\rangle }
\right|^2 N_i(N_j+1) \delta \left[ {\omega  - (E_j - E_i)/\hbar} \right]
\label{eq:sqw}
\end{eqnarray}
The factor $(N_j+1)$ reflects bosonic stimulation by the occupation of the final state. The
elastic term $S_0(q)$ describes coherent elastic scattering or diffraction and is simply the
square of the Fourier transform of the density $ S_0(q) = \left| \langle \rho^{\dag}(q) \rangle
\right|^2 =\left|  \sum  N_i {\left\langle i \right|e^{{\rm i}qr} \left| i \right\rangle
}\right|^2$.

{\it A simple example}. It is instructive to apply this formalism to a system of non-interacting
bosons which has macroscopic occupation in two momentum states with momentum $\pm \hbar k$.  If
the initial state is a Fock state $|+k \rangle^{N_+} |-k \rangle^{N_-}$, we find that, apart from
forward scattering, the dominant term in $S(q, \omega)$ is the bosonically enhanced scattering
between those two (degenerate) states, $S(q,\omega)= \left[ N^2 \delta_{q,0}+ N_+(N_-+1)
\delta_{q,-2k} + N_-(N_+ +1) \delta_{q,2k} \right] \delta(\omega)$ where the Kronecker symbol
$\delta_{q,p}$ implies $q=p$ within the wavevector resolution $\approx 1/L$ of a finite volume
with length $L$.  Alternatively, we can assume the initial state to be a coherent superposition
state $|i \rangle^N$ with the eigenstate $|i \rangle= c_+ |+k \rangle + c_- |-k \rangle$ and
$|c_{\pm}|^2=N_{\pm}/N$ and $N=N_+ + N_-$. Now, the dominant contribution to $S(q,\omega)$ comes
from $S_0(q)=  N^2 \delta_{q,0} + N^2 |c_+| ^2 |c_-|^2 \left[ \delta_{q,2k} + \delta_{q,-2k}
\right]$ which is equivalent to the Fock state case when the difference between $N_{\pm}$ and
$N_{\pm}+1$ can be neglected in the limit of large occupation numbers.

This equivalence between Fock states and coherent superposition states has been extensively
discussed in the context of two interfering Bose-Einstein condensates
\cite{java96phas,nara96,cast97} and also with regard to optical coherences \cite{molm97}.  Those
papers show that, in many situations, a Fock state is equivalent to an ensemble of coherent states
with arbitrary phase. Experimental interrogation determines the phase and reduces the ensemble to a
single coherent state with a phase which will vary from experiment to experiment. For large
occupation numbers, one can therefore regard the Fock state as an initial state which has not yet
``declared its phase'', and, for the convenience of calculations, replace the Fock state by a
coherent superposition state with an arbitrarily chosen phase.

However, on first sight, the physical interpretation is different. In the Fock state formulation,
the enhanced scattering results from a macroscopic occupation number in a single quantum state,
whereas for the coherent superposition state, the scattering is simple diffraction by a
sinusoidally modulated density distribution with an amplitude proportional to $N|c_+ c_-|$. This
density modulation acts as a diffraction grating for incident light or particles and has a
diffraction efficiency proportional to the square of the amplitude. Such a density modulation does
not require bosonic atoms.  It can, for example, be imprinted into thermal or fermionic clouds by
subjecting them to a suitable optical standing wave. The equivalence of these two descriptions
points towards one of the major conclusions of this paper, namely that macroscopic population of
bosonic states is not necessary for enhanced scattering.

The previous discussion assumed scattering between two degenerate momentum states $|\pm k \rangle$.
A simple Gallilean transformation generalizes this to two arbitrary momentum states $|k_{\pm}
\rangle$ with energies $E_{\pm}$.  Now the standing wave moves with a velocity $\hbar (k_+ + k_-
)/2 m$ where $m$ is the mass of the atoms, and the enhanced scattering appears at $\hbar
\omega=\pm(E_+ - E_-)$ instead of at $\omega=0$.

{\it Enhancement of fluctuations}. The general results of statistical physics presented above
emphasize that enhanced scattering {\it must} be related to enhanced density fluctuations.
Therefore, bosonic enhancement of a scattering rate is either due to a density modulation $\langle
\rho(q) \rangle$ (in the coherent superposition description) or due to density fluctuations (in
the Fock state description) --- the latter can be regarded as a density modulation with an unknown
phase. This relation allows a more intuitive answer to the question why there is bosonic
enhancement when two atoms 1 and 2 collide in the presence of a condensate with $N_0$ atoms.  The
standard answer would be that the symmetry of the wavefunction enhances 
the scattering rate into
the condensate and into some other state 3 by a factor of $(N_0 +1)$.  An equivalent answer is
that the condensate interferes with atom 2 (or 1) and creates a density grating with an amplitude
proportional to $N_0^{1/2}$ which diffracts atom 1 (or 2) into state 3. The grating absorbs this
momentum transfer by transferring the atom in state 2 (or 1) into the condensate. Therefore,
bosonic stimulation can be regarded as heterodyne amplification of density fluctuations where the
condensate acts as the local oscillator.

{\it Dicke superradiance}. We now want to establish the connection between bosonic enhancement and
Dicke superradiance.  This will formally introduce the enhancement factor $(N+1)$ for non-bosonic
systems. A system of atoms with $N$ atoms in two states $|\pm \rangle$ is conveniently described
with the formalism introduced by Dicke to discuss superradiance in two-level atoms \cite{dick54}.
It should be emphasized that the only assumption in this treatment is that the $N$ atoms couple
identically to the probe field (the electromagnetic field or some incident particle beam), i.e.,
that they have the same transition frequency and matrix element without any assumption of quantum
statistics.  For example, in magnetic resonance experiments, the Dicke treatment would apply to
different atomic species with the same value of the magnetic moment.

Dicke regarded the two-level atom as a spin 1/2 system and introduced angular momentum quantum
numbers.  In this subspace, a fully symmetric state of $N$ atoms has spin $s=N/2$ and magnetic
quantum number $m=(N_+ -N_-)/2$.  The squared matrix element for the transition $|s, m\pm 1 \rangle
\rightarrow |s, m \rangle$ induced by the ladder operator $S_\mp$ is $(s \pm m +1)(s \mp m)$.
Expressing this by initial occupation numbers $N_\pm$, one obtains $N_\pm (N_\mp +1)$
\cite{sarg74,saku94,wall97mwa} retrieving the formula of bosonic enhancement. The transition rates
are largest for the $N$ particle state with $s=N/2$ which is therefore called the state of maximum
cooperativity.

Such a system will couple to the probe field in a superradiant way (i.e., with an up to $N$ times
enhanced transition rate).  In the Bloch vector picture, its dynamics is described as the
precession of a macroscopic spin vector with length $s=N/2$. This spin vector decays in a time
$1/\Gamma$ where $\Gamma$ is the total (homogeneous and inhomogeneous) linewidth of the transition
$|+ \rangle \rightarrow |- \rangle$. Collective superradiant behaviour can only be observed at
times shorter than $1/\Gamma$.

{\it Application to matter wave gratings}. Dicke's formalism is usually applied to one-photon
transitions between internal states, but here we use it to discuss scattering, i.e.\ a two-photon
transition between two momentum states $|k_{\pm} \rangle$.  Let's first assume that we have an
ideal Bose-Einstein condensate in the $k=0$ momentum state. Light scattering between momentum
states $k=0$ and $k=q$ has an infinite coherence time for a non-interacting condensate of infinite
size (Fig. 1a).  For a thermal (non-degenerate) cloud of atoms with thermal momentum spread $\hbar
k_{\rm th} \ll \hbar q$ the transition frequencies for the transfer of momentum $\hbar q$ are
Doppler broadened by $\Gamma=\hbar k_{\rm th} q/m$. For times shorter than $1/\Gamma$ the system
will behave collectively like the Bose condensed system, i.e.\ a probe beam would induce
transitions between the $k=0$ and $k=q$ momentum states at a rate proportional to $N_{k=0}
(N_{k=q} + 1)$.  The same argument applies to a Fermi degenerate cloud by replacing $k_{\rm th}$
with the Fermi wavevector $k_F$ (Fig. 1b).  Due to the assumption $\hbar k_F \ll \hbar q$, Pauli
blocking due to scattering into already occupied states is absent. If this assumption is not made,
a part of the cloud becomes inactive, and our discussion would apply only to the atoms near the
Fermi surface.

The previous paragraph generalized the bosonic {\it Fock state} ensemble to non-degenerate and
fermionic clouds.  We now come back to the {\it coherent superposition} state. For bosons, it can
be produced from a Bose-Einstein condensate in the $\vec{k}=\vec{0}$ state by applying a (so-called
Bragg) pulse of two laser beams which differ in wavevector by $\vec{q}$ and in frequency by the
recoil frequency $\hbar^2 q^2/2m$. Those beams resonantly drive the transition between momentum
states $\vec{k}=\vec{0}$ and $\vec{k}=\vec{q}$ \cite{kozu99bragg,sten99brag} and prepare the
superposition state discussed above. Similarly, in a thermal (or fermionic) cloud, the Bragg pulse
creates a modulated density distribution with wavelength $2 \pi/q$ which has the same contrast as
in the bosonic case and will diffract light or atoms at the same rate. However, due to the thermal
motion with velocity $\hbar k_{\rm th}/m$, this grating decays during a time $m/\hbar k_{\rm th} q
= 1/ \Gamma$ (and similar for the fermionic case). Thus the Dicke picture and the diffraction
picture agree.

\begin{figure}[htbf]
\epsfxsize=84 mm \centerline{\epsfbox{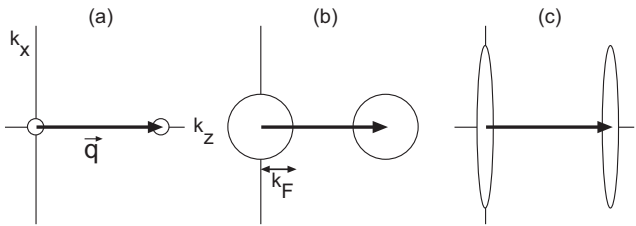}}\vspace{0.3cm}
    \caption[Title]{Momentum transfer $\vec{q}$ (a) to a Bose-Einstein condensate, (b) to a Fermi
sea, and (c) to a momentum squeezed degenerate Fermi cloud.  Shown are 
the populated states vs. the
$k$-vector. The momentum spread $k_F$ of the Fermi sea introduces Doppler broadening of the
transition and a finite coherence time, whereas the coherence time in (a) and (c) is infinite.
    \label{fig:Momentum_transfer}}
\end{figure}

{\it Coherence time}. The Doppler broadening discussed above seems to imply a fundamental limit
to the coherence time of a Fermi system.  However, at least in principle, one can prepare a Fermi
system with infinite coherence time by starting out with a cloud which is in a single momentum
state along the $\hat z$ axis, but occupies many momentum states along $\hat x$ and $\hat y$.
With a Bragg pulse transferring momentum $q \hat z$, one can prepare a system which shows
collective behavior for scattering particles or light with momentum transfer $q \hat z$ with an
infinite coherence time (Fig. 1c). Therefore, there is no direct connection between a long
coherence time and a high phase-space density. In this ensemble, the scattering is between the
states $|k_z=0 \rangle \otimes |k_x, k_y \rangle$ and $|k_z=q \rangle \otimes |k_x, k_y \rangle$.
Therefore, we have enhanced scattering into the $|k_z=q \rangle$ quantum state, but the atoms may
differ in other quantum numbers. What matters is only the symmetrization of the many-body
wavefunction along $\hat z$.  The other quantum numbers ensure that there is no conflict with the
Pauli blocking for fermionic systems.  This is analogous to the separation of electronic
wavefunctions into a symmetric part (e.g. the spin part) and an antisymmetric part (e.g. the
spatial part) where the coupling to an external field (e.g. electron spin resonance experiment)
only depends on the symmetric part.

{\it Experiments}. The experiments both on superradiance \cite{inou99super} and four-wave mixing
\cite{deng99} in Bose-Einstein condensates have in common that a matter wave grating formed by two
macroscopically occupied momentum states is probed, either by light or by atoms. Both experiments
create the coherent superposition state discussed above using a Bragg pulse. In the limit of low
intensity of the probe beam, the scattering is independent of the nature of the probe particles
--- one could have used any kind of radiation, bosons or fermions \cite{vill00fwm}. The bosonic stimulation
observed in both experiments demonstrates the dynamic nature of the matter wave grating.  Each
time, a particle or photon is diffracted, the amplitude of the grating grows.

In practice, it is difficult or impossible to carry out these experiments with fermions or thermal
atoms.   When we observed superradiance of a condensate, we couldn't observe similar behaviour
above the BEC transition temperature since the threshold laser intensity for superradiant gain is
several orders of magnitude higher (see Ref.\ \cite{inou99super}) for details).  Furthermore, the
superradiance may be suppressed by heating or other decoherence processes. The shorter coherence
time for non-BEC samples should be even more crucial for the four-wave mixing experiment where the
matter wave grating is probed by very slow atoms which have a long transit time of about 1 ms
through the sample. Another concern are incoherent processes which accompany the stimulated
processes discussed so far. Since the incoherent processes scale linearly with the number of
atoms, whereas the stimulated process is proportional to $N^2$, there is in principle always a
regime where the stimulated process dominates \cite{fn:fermi}.

{\it Discussion}. Coming back to the initial question:  Is matter wave amplification possible for
fermions?  The answer is yes, if the system is prepared in a 
cooperative state and the amplification
is faster than the coherence time.  However, this amplification does not pile up atoms in a single
quantum state, but rather in states which are in the same (or approximately the same) momentum
state along $\hat z$, but differ in other quantum  numbers. Therefore, this amplification can be
regarded as amplification of a density modulation or as amplification of spatial bunching of atoms.

The phase-coherent matter wave amplification for fermions would start with a short Bragg pulse
which puts some of the atoms into a recoil state which is then amplified.  This superposition of
two momentum states creates a matter wave grating.  This can be regarded as the interference
pattern of each atom with itself with all the individual interference patterns being exactly in
phase. Matter wave amplification occurs when a single laser beam is diffracted off this grating
increasing the amplitude of each atom to be in the recoiling state. Therefore, the matter wave
amplification scheme of Refs.\ \cite{inou99mwa,kozu99amp} would work for fermions, provided the
whole process can be done in the short coherence time of the fermionic matter wave grating.

The major difference between bosonic and fermionic system is that a bosonic system with two
macroscopically occupied quantum states is {\it always} in a fully symmetric 
and maximally cooperative
state.  In other words, if two independent Bose condensates cross each other, there is always a
macroscopic interference pattern (as observed experimentally \cite{andr97int}), which is reflected
in $S(q, \omega)$ being proportional to $N^2$ (or to $N_+ N_-$, to be more precise). It is this
density modulation which can be amplified by the dynamic diffraction discussed in this paper.  If
two beams of fermions overlap, there is no macroscopic interference, unless the two beams were
prepared in a symmetric way, e.g. by generating one of the beams by a Bragg pulse from the other
one.

Our discussion of scattering without change of the internal state can be generalized.  For example,
if atoms scatter into the condensate through a spinflip process, the density grating has to be
replaced by a polarization or coherence grating.  Such gratings were experimentally studied for
laser-cooled atoms in Ref.\ \cite{kuma98}.

This paper has focused on bosonically enhanced {\it scattering}.  Similarly, bosonic enhancement
of spontaneous emission depends only on a cooperative initial state and not directly on quantum
statistics. For scattering, the relevant coupling strength are the density fluctuations.  For
spontaneous emission, it is the electric dipole moment.  Both are enhanced by the presence of a
Bose condensate, in the latter case because the excited atom corresponds to a Dicke vector of spin
$s=N/2, m=-N/2+1/2$ which couples more strongly to the vacuum fluctuations of the electromagnetic field
than an individual atom. Alternatively, the enhanced spontaneous emission 
can be regarded as the constructive
interference of an ``emitted'' ground state atom with the macroscopic ground state matter
wave.  This picture is analogous to the semi-classical interpretation of stimulated emission of
light.  Ref.\ \cite{sarg74} shows that bosonic stimulation of photons is due to the constructive
interference of the emission of a classical oscillating dipole with the incident field in the
forward direction.

In conclusion, we have shown that bosonically enhanced scattering is related to density
fluctuations and matter wave gratings.  The analogy with Dicke superradiance emphasizes that
matter wave amplification and atomic four-wave mixing are possible for fermionic or non-degenerate
samples. Bosonic quantum-degeneracy is sufficient, but not necessary for such enhanced
scattering.  It represents only one special way to prepare a system in a cooperative state which
shows coherent and collective behavior.

We are grateful to A. Aspect, C. Cohen-Tannoudji, K. Helmerson, D. Kleppner, P. Meystre, W.D.
Phillips, D.E. Pritchard, S. Rolston, M.O. Scully, D.M. Stamper-Kurn, and E. Wright for helpful
discussion, and to A. G\"orlitz, C. Raman, A.P. Chikkatur and S. Gupta for valuable comments on the
manuscript. This work was supported by NSF, ONR, ARO, NASA, and the David and Lucile Packard
Foundation.


\begin{thebibliography}{10}
\bibitem{fn:fermi}
In the case of a fermionic sample, the density of particles can only be increased by increasing
the momentum spread of the sample.  This can increase the incoherent ``background'' for four-wave
mixing.  For the ensemble in Fig. 1c an increase in $N$ at constant volume requires the transverse
velocity spread to increase as $N^{1/2}$. Therefore, for large $N$,the incoherent elastic
scattering rate increases as $N^{3/2}$, still more slowly than the stimulated scattering.
\end{thebibliography}

\end{document}